# Non-invasive, near-field terahertz imaging of hidden objects using a single pixel detector

R. I. Stantchev[1*], B. Sun[2], S. M. Hornett[1], P. A. Hobson[1,3], G. M. Gibson[2], M. J. Padgett[2], and E. Hendry[1]

[1]School of Physics, University of Exeter, Stocker Road, Exeter EX4 4QL, UK
[2]SUPA, School of Physics and Astronomy, University of Glasgow, Glasgow, G12 8QQ, UK
[3]QinetiQ Limited, Cody Technology Park, Ively Road, Farnborough, GU14 0LX, UK

**Abstract:**

**Terahertz (THz) imaging can see through otherwise opaque materials. However, due to the long wavelengths of THz radiation (λ=300μm at 1THz), far-field THz imaging techniques suffer low resolution compared to optical systems. In this work we demonstrate non-invasive near-field THz imaging with sub-wavelength resolution. We project a time-varying, intense (>100μJ/cm2) optical pattern onto a silicon wafer which spatially modulates the transmission of synchronous pulse of THz radiation. An unknown object is placed on the hidden-side of the silicon and the far-field THz transmission corresponding to each mask is recorded by a single element detector. Knowledge of the patterns and the corresponding detector signal are combined to give an image of the object. Using this technique we image a printed circuit board on the underside of a 115μm thick silicon wafer with ~100μm (λ/4) resolution. With sub-wavelength resolution and the inherent sensitivity to local conductivity, we show that it is possible to detect fissures in the circuitry wiring of a few microns in size. THz imaging systems of this type will have other uses too, where non-invasive measurement or imaging of concealed structures is necessary, such as in semiconductor manufacturing or in bio-imaging.**

Due to its unique properties, imaging and analysis with THz radiation has attracted a lot of attention in recent years [1, 2, 3]. For example, the transparency of most non-conductive materials in the THz range is extremely useful for systems inspection [4], and allows THz measurements to uncover the material composition and substructure of paintings, murals or frescoes [5]. The non-ionizing photon energies are of interest to medical imaging applications, where resonant Debye relaxation of small molecules such as water gives rise to useful image contrast [6, 7]. However, unlike the optical domain, the terahertz regime is plagued by a lack of materials suitable for the construction of cheap and reliable focal plane imaging arrays, giving the rise to the 'THz gap' [3]. Further more, due to the long wavelengths (0.15-1.5mm), THz imaging is severely handicapped by the diffraction limit, restricting biological imaging, for example, to large structures such as organs [8, 9]. There has therefore been tremendous effort to develop sub-wavelength THz imaging techniques. These typically rely on some form of raster scanning of a local modulator [10-14], or of the THz detector itself [15-21], in the near field. Undoubtedly the most impressive THz imaging resolution has been achieved using tip scattering of near fields, where imaging of single nanoparticles is possible [19]. While these straightforward scanning approaches have yielded tremendous improvements in terms of resolution, they are also inherently slow, often invasive, and generally more suited to solid state, conductive samples with well-defined interfaces [18, 19].

In recent years, alternative imaging approaches have emerged which use spatially controlled light, where the reflected, transmitted or scattered radiation is recorded using a single element detector [22-26]. These approaches have both practical and economic advantages by completely dispensing with the need for slow mechanical scanning or expensive multip-ixel detectors. Further, these alternative imaging methods are compatible with compressed sensing [27] where one takes an $N$ pixel image with $M<N$ measurements, something unattainable by the imaging approaches of Refs. [10-21]. Such single element detection schemes have recently been demonstrated for far-field, diffraction-limited THz imaging with a typical resolution of ~1mm [28-34], and it has been speculated that similar concepts could in principal be developed for efficient near field THz imaging [35].

In this letter, we explicitly demonstrate near-field THz imaging using a single-element THz detector that is able to detect micron sized fissures in a circuit board hidden on the underside of a silicon wafer. Our THz source is spatially modulated in the near field by a second optical source projected simultaneously onto a thin photoconductive modulator, which is itself placed in the near field of an object. Our technique combines many of the advantages of traditional THz imaging (such as transparency to non-conducting materials) with sub-wavelength spatial resolution. Since the spatial resolution of the imaging is fundamentally determined by the optical source, this approach retains the tantalizing prospect of non-invasive THz imaging with without constraint of the diffraction limit.

Our imaging set up is illustrated in figure 1 (a more detailed schematic is shown in figure S1 in the supplementary information, section §1). We use a time domain measurement of a broadband THz pulse (0.2 - 2THz), Figs. 2**a**-**b**. To spatially modulate our THz beam, we shine a coincident 800nm, 100fs pump pulse onto a highly resistive silicon wafer (1000Ω·cm, 115μm thick). The pump pulse itself is structured into binary spatial intensity patterns by a standard digital micromirror device [36]. When these patterns are projected onto the silicon wafer, the photoexcited regions are rendered conductive (see supplementary info section §2) and thus also opaque to the coincident THz radiation [37]. Moreover, because we record the THz transmission immediately following photoexcitation, before processes such as electron diffusion take place (see methods), the spatial pattern encoded in the 800nm pulse is directly transferred to the THz pulse without smearing or broadening of spatial features. The patterned THz pulse then propagates through ~115μm of silicon before interacting with a sample positioned on the hidden side of the wafer, after which we record the far-field transmission (see methods section). By spatially encoding a beam of THz radiation with binary intensity patterns, an image can be formed by analyzing the THz radiation transmitted or scattered by an object using a single pixel detector [26, 38] (see methods and supplementary info section §3). Further, since the distance travelled by the THz pulse before encountering the object is relatively small compared to the wavelength, we can record an image before far-field Fraunhofer diffraction occurs. In Figure 2**d** we show a measured THz image of the metallic cartwheel shown in Fig. 2**c**. A cartwheel is chosen as an object here as it contains ever increasing spatial frequencies towards the center of the wheel, allowing us to estimate the resolution of our image. The white arrows in figure 2**d** indicate an estimate of the imaging resolution, evaluated as the minimal distance between the arms of the cartwheel for which the image contrast is not diminished due to diffraction. From this, we find a resolution of 103(±7)μm, significantly smaller than the 375μm peak wavelength of our THz pulse (see fig. 2**b**). Using scalar near field diffraction theory [39], we can explicitly show that the resolution expected for our measurement is ~95μm (see figure S4 in the supplementary information section §4), in reasonable agreement with our experimental estimate. It should be noted that this resolution is by no means a fundamental limit, and is determined primarily by the finite thickness of the photomodulator (the silicon wafer). The approach outlined above combines many of the advantages of traditional THz imaging, such as transparency to non-conducting materials [40], with sub-wave-length spatial resolution provided by the optical modulation. This makes it particularly suitable for imaging small structures buried beneath optically opaque, non-conducting materials. We now demonstrate a potential application: imaging of a printed circuit board, hidden on the underside of a silicon wafer (see Fig. 3**a** for design with dimensions).

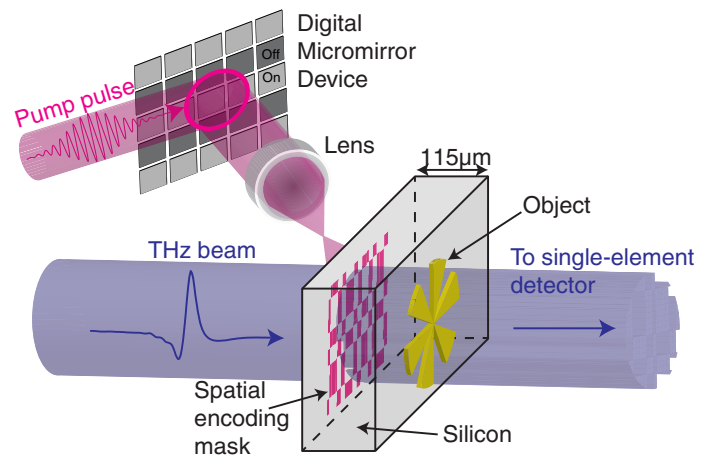

**Figure 1: Schematic illustration of near field, single pixel THz imaging.** The imaging scheme: an optical pump pulse is spatially modulated and used to photoexcite a semiconducting wafer, which transfers the spatial encoding mask onto a coincident THz pulse. The subsequent THz pulse is then passed through an object onto a single pixel detector.

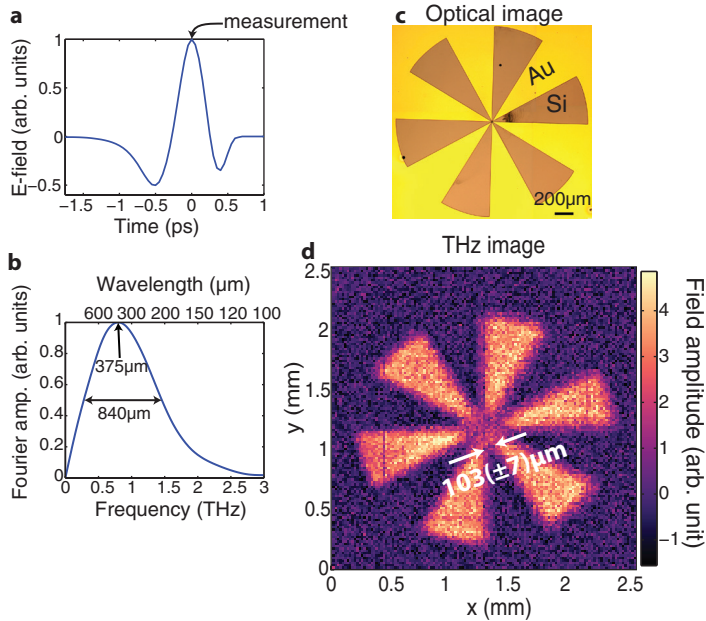

**Figure 2: Pulsed THz Imaging. a,** The electric field of our THz pulse recorded in the time domain using electro optic sampling. The arrow shows the measurement point, at the peak of the THz field, for which images are recorded. **b,** Normalized Fourier transform of our THz pulse. The central wavelength is approximately 375μm, with a full width half maximum of 840μm. **c,** An optical image of a resolution test target. Au marks the regions spanned by the gold film, while the regions marked Si show the exposed silicon wafer. **d,** 128×128 THz image of the resolution test target in **c**, obtained via a full set of Hadamard masks. The pixels are 20μm in size. The arrows indicate the imaging resolution, evaluated as the maximal distance between the arms of the cartwheel for which the image contrast is diminished due to diffraction.

We start by addressing the strategies by which we can record the data necessary to reform an image. The most straightforward way to obtain an image is to illuminate each pixel sequentially, as in raster scanning, and record the THz signal for each pixel. The resulting image using this approach, shown in figure 3**b**, contains a lot of noise arising from both our THz generation and detection. However, more complex multi-aperture masking schemes can be devised using binary matrices for illumination. These multi-pixel illumination schemes offer the advantage of minimizing the effect of detector noise by using more light in each measurement [26]. In terms of the choice of masks, we can employ random binary masks, as used in compressed sensing [27]. However, provided one has a stable light source, the best results are usually given by masks derived from Hadamard matrices [41], namely the masks form an orthonormal set that minimizes the mean squared error in each image pixel [26]. In figures 3**b-d** we compare images formed using raster, random and Hadamard masks measured under identical conditions. While the multi-pixel schemes offer a clear advantage over sequential raster scanning, we also observe a clear superiority of Hadamard over random masks when we use a simple image reconstruction algorithm. To construct our images in Figs. 3**b-d**, we sum the masks with each one weighted by the detector readout for that mask. This algorithm is advantageous due its fast computation (<100ms), and for the Hadamard case it recovers the exact solution. A more in-depth comparison for various image sizes is carried out in supplementary info section §5.

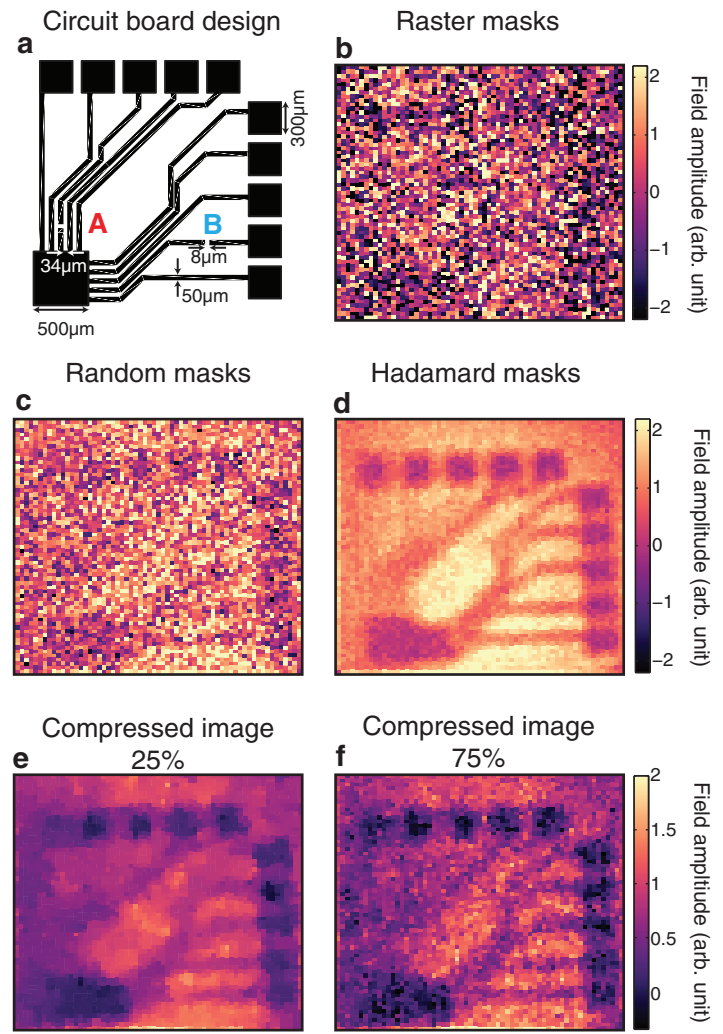

**Figure 3: Hadamard vs Random vs Raster imaging. a,** The circuit board design, where black indicates conducting, metallic regions. The individual wires are 50μm in width, and 8μm breaks have been introduced at points marked by the letters A and B. **b,** Image acquired using raster scanning of a single opaque pixel. **c** and **d,** Comparison of the same image acquired with a full set of masks derived from random and Hadamard matrices, respectively. **e** and **f,** Compressed images obtained via random masks where the number of measurements is 25% in **e** and 75% in **f** of the total number of pixels (we use a total variation minimization image recovery algorithm, see supplementary information section §6 for more details). In all images the THz electric field is polarized horizontally and number of pixels is 64×64 with 40μm pixels. The signal acquisition time for a single measurement is 500ms. Due to the the considerably larger noise in the measurement, we have scaled the image in **b** by 0.25 in order to use the same colour scale as in parts **c** and **d**.

The approach described above requires $N$ measurements to obtain an $N$ pixel image. However, more sophisticated image recovery algorithms have been developed by the field of compressed sensing [27] allowing one to image using $M<N$ measurements. For this, we use a total variation minimization algorithm to recover our compressed images (see supplementary info section §6 for details). In figures 3**e** and **f**, we show the effect of decreasing the number of measurements when sampling our circuit board with random masks. The structure of the circuit board can be observed even when only 25% measurements are used (Fig. 3**e**). However, some of the fine detail is missing due to under-sampling. As the number of measurements are increased, the images begin to resemble the image measured using Hadamard masks in Fig. 3**d**. However, here the image quality is primarily determined by level of post-processing that one performs, as discussed in supplementary info section §6. While compressive imaging can cut down measurement time, it does so at the cost of post-processing and image detail. In the remainder below, we therefore discuss results obtained with Hadamard imaging.

Our THz source is linearly polarized, thus we can expect effects due to the polarization boundary conditions. In particular, the electric field component parallel to the interface of a good conductor must approach zero. These effects are particularly prominent due to the subwavelength nature of the conducting features in our circuit board (the thickness of the conducting wires is 50µm and the separation between the individual wires at some locations is ~30µm). In Fig. 4**a** we show a THz image of the circuit board in 3**a** as measured with vertical THz polarization. We see that the subwavelength conducting wires are more clearly observed when the THz radiation is parallel to the wires. The biggest difference is seen in the conducting tracks emerging from the large 500µm square in the bottom left corner, Figs 4**b** and **c**. Here, the small separation of the wires resembles a wire grid polarizer, with the transmission lowest (and image contrast highest) when the polarization is parallel to the wires.

While such effects may be seen as a disadvantage, for example by limiting the resolving capacity for some metallic features, we discuss below how one can use polarization sensitivity to our advantage by employing it to detect very subwavelength features. To this end, we have introduced very small ($\leq$ 8µm) fissures in wires at two points marked by A and B in circuit diagram in Fig. 3**a**. In Figs. 4**b** and 4**e**, these fissures appear as marked increases in the THz transmission amplitudes at the points identified by dashed circles. Note that, in order to better distinguish these subwavelength features, Figs 4**b-c**

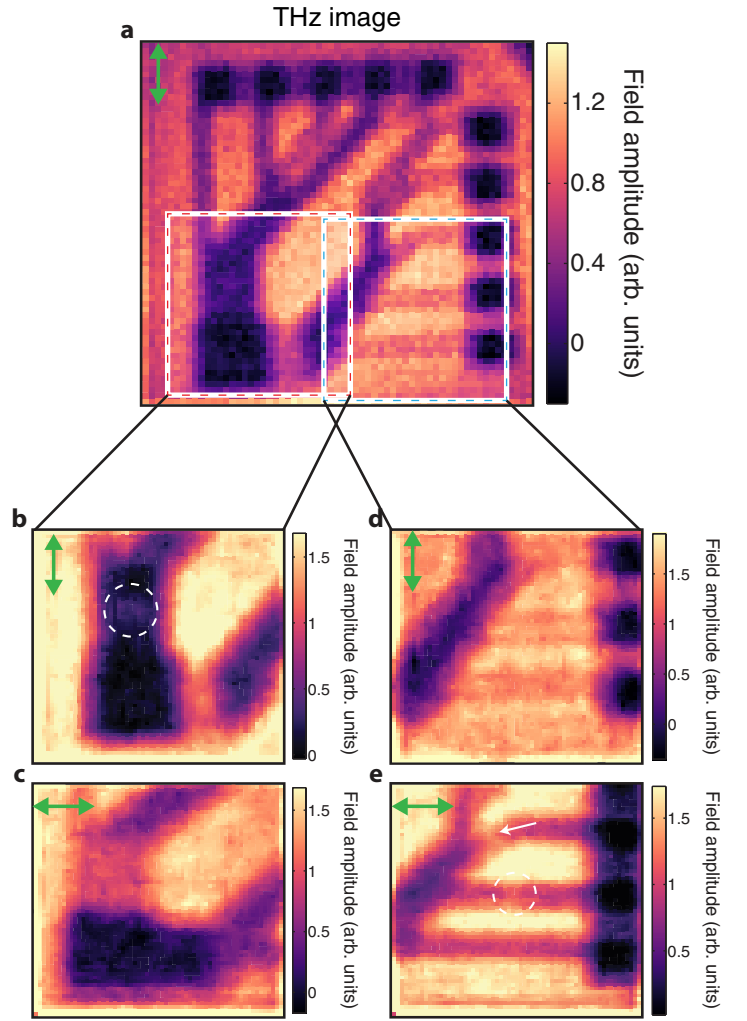

**THz image**

**Figure 4: Imaged polarization effects. a,** 64×64 THz image of circuit board in Fig 3**a** with vertical polarization. Pixels are 40µm. We see that the contrast of each of the individual wires in the circuit depends on the THz polarization, with the highest contrast seen for polarization parallel to the wires. **b-e,** 64×64 images of the square regions in part **a**. Polarization is shown by the green arrow in the top left corner of each picture. Pixel size is 20µm, and images have been denoised using the algorithm described in supplementary section §7. We see that the very subwavelength wiring breaks (marked by circles in parts **b** and **e**) give rise to transmissive regions in the THz image when the THz polarization is parallel to the wire. In **e,** the diagonally orientated wire (indicated by the white arrow) also shows low contrast. Every image has been obtained via a full set of Hadamard masks.

have been denoised using the algorithm outlined in ref. [42] (see supplementary information section §7 for a comparison of different filtering approaches). It should be noted that the observed increase in THz transmission is considerably larger than one would expect solely from the reduced coverage of gold, and arises from a relaxation in the parallel field boundary condition due to the presence of the fissure. The subwavelength fissures are not visible when the polarization is perpendicular to the wires, as shown in Figs. 4**c** and **d**. One can therefore not only identify the orientation of the wiring using our approach, but also detect extremely subwavelength defects in circuitry hidden beneath optically opaque silicon.

In conclusion, we have demonstrated non-invasive, subwavelength THz imaging using a single-pixel detector. With subwavelength ($\sim\lambda/4$) resolution, we demonstrate a proof of principal application where we image a printed circuit board on the underside of a 115µm thick silicon wafer, and show how polarization sensitivity can be utilized to detect sub-wavelength breaks in its thin conducting tracks. It is interesting to note that the fundamental limit for resolution in this approach is the optical diffraction limit. With potential for significantly greater resolution and faster acquisition rates, using thinner and faster (direct band gap) photomodulators, the imaging approach discussed here gives rise to several intriguing prospects, such as imaging of conducting channels in biological systems.

## Methods

An amplified 800nm (100fs) Ti-Sapphire femtosecond laser running at a repetition rate of 1050Hz is used to power a THz time-domain spectrometer. The THz pulses are generated and detected, using optical rectification and electro-optic sampling respectively [43, 44], in ZnTe crystals. The femtosecond pulses also provide an optical modulation beam with fluences of 107μJ/cm². The pump pulse is spatially-modulated via a digital micromirror device (DLP3000 with the DLP Lightcrafter from Texas Instruments with the original LED illumination optics removed to allow direct access to the micromirrors) and a single lens (focal length 75mm) so as to project a binary pattern on the surface of a high resistivity silicon wafer (1000Ω·cm, 115μm thick). The modulation pulse is coordinated with a collimated THz pulse in both space and time to arrive coincident at the front interface of the silicon wafer. We measure the THz transmission within a ~5ps window after photoexcitation, hence carrier diffusion can be neglected, as shown in supplementary section §2, and thus our spatial pattern is directly imprinted onto our THz beam. The ultrafast synchronization between the optical pump and the THz pulse is one of the key developments from the work in ref. [21] which has allowed us to access the sub wavelength imaging regime. For our images, we record the peak of the THz pulse (Fig. 2**a**) transmitted through a sample, placed directly after the silicon modulator, in the far field. This gives a spectrally averaged weight to our measurements, centered around the peak spectral wavelength of 375μm (see Fig. 2**b**). In principle, one can take images at all temporal points of the THz pulse and thus obtain full spectral information.

To obtain an image, we record a total of $N$ THz transmission measurements for $N$ distinct spatial encoding masks. In matrix notation this can be represented as $\Phi = W\Psi$ where $\Phi$ is a vector of the sequential measurements made, $W$ is a measurement matrix where the $(i,j)$th entry determines the value of the $j$th mask pixel in the $i$th measurement and $\Psi$ is an $N$-pixel image of the object. The image can be obtained by inverse matrix multiplication: $\Psi = W^{-1}\Phi$ (see supplementary info section §3 for more information), but other methods exist if the matrix can not be inverted [27]. Our binary transmission masks have either opaque or transmissive pixels ie. are described by 1s and 0s. In order to preserve the orthonormallity of Hadamard matrices, which are composed of +1s and -1s, we carry out sequential measurements of a mask directly followed by its inverse and record the difference in THz transmission for these measurements via a lock-in amplifier. This differential measurement is described by matrices with elements of +1 and -1 as is outlined in Ref. [38], and in supplementary Fig. S3 we explicitly show that this results in a fourfold improvement in our images. The signal acquisition time for each mask and its inverse is 50ms. It should be noted that the fundamental switching rate is determined in our measurements by the low repetition rate of the laser (1050 Hz). In principal, using a higher repetition rate laser could greatly increase the acquisition rate, which will be ultimately limited by the recovery time of the (MHz) photomodulator.

Samples are fabricated on the rear interface of the silicon wafer using 250nm gold films deposited via thermal evaporation. A 5nm layer of chrome acts as the adhesion agent between the silicon and the gold. Image patterns are wet etched in the gold layer using electron beam lithography. At THz frequencies, the response of good conductors such as gold is essentially dispersionless. In order to investigate polarization effects in the experiment samples are rotated, while the horizontal THz polarization remains fixed.

### Author Contributions

E. H. and M. J. P. conceived the idea. R. I. S. and E. H. designed the experiment, and R. I. S., B. S., and G. M. G. setup the experiment and micromirror array controls. S. M. H. manufactured all samples. R. I. S. performed the experiments. R. I. S. and B. S. performed the data analysis. M. J. P., B. S. and G. M. G. designed the denoising algorithm. R. I. S. and E. H. wrote the manuscript and all other authors provided editorial input.

### Acknowledgement
The research presented in this work was funded by QinetiQ & EPSRC under iCase award 12440575 and grant number EP/K041215/1.

# Supplementary Information of:

# "Non-invasive, near-field terahertz imaging of hidden objects using a single pixel detector"

R. I. Stantchev[1*], B. Sun[2], S. M. Hornett[1], P. A. Hobson[1,3], G. M. Gibson[2], M. J. Padgett[2], and E. Hendry[1]

[1]School of Physics, University of Exeter, Stocker Road, Exeter EX4 4QL, UK
[2]SUPA, School of Physics and Astronomy, University of Glasgow, Glasgow, G12 8QQ, UK
[3]QinetiQ Limited, Cody Technology Park, Ively Road, Farnborough, GU14 0LX, UK

*ris202@exeter.ac.uk

## §1 Experimental Schematics

As stated in the main text, our THz pulses are generated and detected within a Terahertz time-domain spectrometer (THz-TDS) [1s, 2s, 3s]. The fundamental layout of the setup is shown in supplementary figure S1. In its essence, a beam of femtosecond optical pulses is split into three beams: generation, detection and excitation. The first is used to generate a picosecond THz pulse, we use optical rectification in a ZnTe crystal [4s, 5s], which then passes through the sample under investigation. Our THz beam is collimated and collected by 90° off-axis parabolic mirrors made from aluminum, both with a 2.5cm focal length and 2.54cm diameter. The second beam is used to detect the time profile of the THz waveform. This is achieved by temporally overlapping the much longer THz pulse with the very short detection pulse. The difference in pulse durations, allows one to discretely sample the terahertz temporal profile by varying the path lengths with an optical delay line (typical THz transient and detection pulse envelope shown in figure S2**a**). The electric field amplitude is extracted via electro-optic sampling in a ZnTe crystal [6s, 7s]. The third beam is used to photoexcite the sample. Additionally, our excitation beam is spatially modulated via a digital micromirror device (DMD) and a lens so as to project any binary pattern onto our sample.

We use a single +7.5cm focal length lens to project the binary patterns from the DMD onto our silicon, with a magnification of -0.66. Our DMD (DLP3000 used on a DLP Lightcrafter from Texas Instruments) has square mirrors of size 15.2μm, hence we are limited to projecting squares of size ~10μm. Using the Rayleigh lens formula, $\theta=1.22\lambda/D$, our 800nm pulses have a diffraction limited resolution of ~5.4μm at our imaging plane.

## §2 The silicon photomodulator

We photoexcite electron-hole pairs in silicon using ultrafast (90fs) pulses with a wavelength of 800nm. The photoexcited dielectric function of silicon can be described by the Drude model [8s, 9s]

$$\varepsilon(\omega) = \varepsilon_b - \frac{\omega_p^2}{\omega^2 + i\omega/\tau_c}, \quad (1)$$

where $\varepsilon_b$=11.7 is the background dielectric permittivity due to the bound electrons, $\tau_c$ is the average collision time with typical values $\tau_c \approx 0.5$ps for the excitation energies used here [9s], $\omega_p$ is the plasma frequency defined as $\omega_p^2 = n_c e^2/\varepsilon_0 m^*$ with $e$ being the electron charge, $\varepsilon_0$ free space permittivity and $m^*$=0.26$m_e$(0.37$m_e$) the effective mass of electrons (holes) [10s].

The primary modulating parameter in the equation above is the density of carriers, $n_c$. Carriers are generated by pulses running at a repetition period of 1ms, considerably longer than the silicon carrier lifetimes of ~25μs [11s], thus we neglect sample heating. Moreover, since the THz pulse arrives ~5ps after photoexcitation, as shown in supplementary Fig. S2**c**, we can also neglect carrier recombination and diffusion. The key variable to determine is therefore the mean carrier-carrier distance. For the 5ps in-between photoexcitation and the arrival of the THz pulse, we can calculate the mean square displacement of carriers, $\langle x^2 \rangle = 6Dt$, where $D$ is the diffusion coefficient of our electron (hole) charge carriers. We use the Einstein–Smoluchowski relation, $D=\mu_q k_B T/q$, where $\mu_q$ is the mobility of charge carriers given by $\mu_q=q\tau_c/m^*$ [12s], to obtain mean displacements of $\sqrt{\langle x^2 \rangle}$=506nm(425nm) for our photo-electrons (holes). Since the diffusion lengths are considerably smaller than the penetration depth of the photoexcitation light (~11μm [13s]), we can neglect carrier diffusion from our considerations. The carrier density is then given directly by the absorbed photon density. This gives $n_c = 2P\rho/d\hbar\omega_l$ where $P$ is our pulse energy per unit area (104μJ/cm²), $\rho$=0.7 is the Fresnel transmittance of Si at our excitation wavelengths, $\hbar\omega_l$ is the photon energies of the pump light, $d$ is the penetration depth ($d\approx11\mu$m [13s] for our wavelengths) and the factor of 2 accounts for the electron-hole pair.

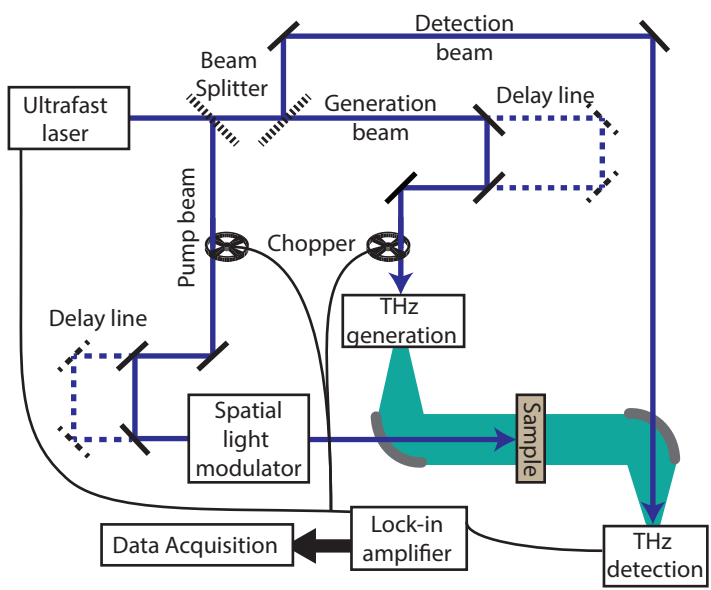

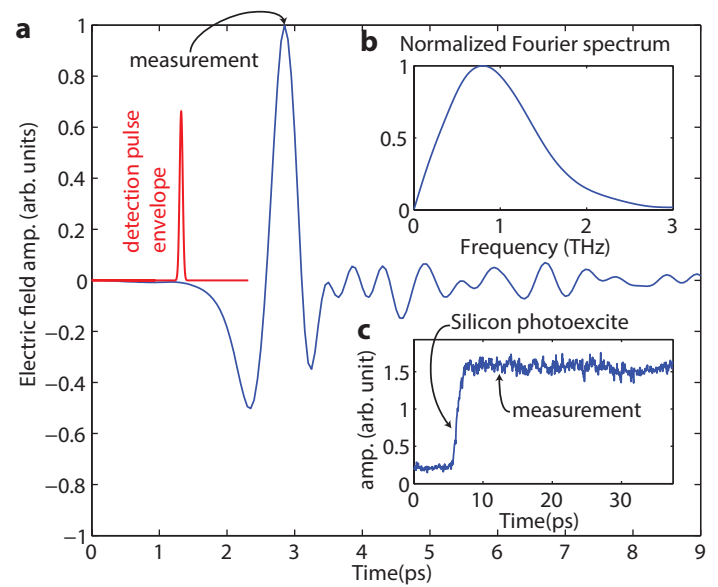

**Supplementary figure S2: Terahertz spectroscopy. a,** Blue: typical THz pulse detected by our system. Red: envelope of detection pulse used to discretely sample the THz waveform. Measurement arrow points to the THz amplitude we detect for each individual mask readout. Oscillations after main THz pulse are due to water vapour in the background environment. **b,** Fourier spectrum of the THz pulse with water vapour oscillations zeroed out. **c,** Modulated THz transmission due to the photoexcitation of a silicon wafer. Measurement arrow indicates where the measurements are made with respect to silicon photoexcitation (~5ps after).

**Supplementary figure S1: Schematic of time domain terahertz spectrometer.** A beam of ultrashort optical pulses leave an ultrafast laser. The beam is split into three beams: generation, detection and excitation. A chopper is placed in the detection or generation beams, depending on the needs of the experiment. Parabolic mirrors are used to collect and collimate the THz radiation.

Entering these values, we obtain a plasma frequency of 81THz with $\varepsilon(1\text{THz}) = -138 + 48i$ for our photoexcited silicon. In other words, we generate a THz material with a negative real and positive imaginary part to the dielectric function, the characteristics of a lossy conductor.

### §3 Single pixel detector imaging theory

The THz regime has the problem that detector arrays are difficult and expensive to manufacture [21s], hence the need for imaging with single pixel detectors. The disadvantage to single pixel imaging is that it typically requires longer acquisition time compared to focal plane imaging arrays, due to the measurements being taken sequentially rather than in parallel. Here we are concerned with obtaining the spatial transmission function of some object using a single pixel THz detector. The simplest solution is to raster scan a single aperture to obtain the transmissivity pixel by pixel. However, if this aperture is made smaller and smaller, the detected signal is reduced and eventually one will run into detector noise.

One could sample more than one aperture simultaneously to increase the detected signal level in order to overcome detector noise, an idea which seems to first originate in 1935 with Yates [22s]. To do this, however, introduces extra calculation difficulties as the measured intensity is due to the sum of the scanning apertures. Therefore, the locations of the scanning apertures in each measurement must form a set of simultaneous equations which can be solved exactly for the individual pixels of the object's transmission function. The information regarding the location of the scanning apertures is held in mask configurations.

We now consider the construction of an $N$-pixel image $\Psi$. Our $i^{th}$ measurement, $\phi_i$, is the dot product of our object transmission function and the $i^{th}$ mask configuration, mathematically expressed as

$$\phi_i = \sum_{j=1}^{N} w_{ij}\psi_j, \tag{2}$$

where $w_{ij}$ holds the spatial information of the $i^{th}$ mask and $\psi_j$ is the $j^{th}$ pixel of the image. As stated in the main text, this can be represented by the matrix equation $\Phi = W\Psi$, where the rows of $W$ are reformatted into the projected masks. For invertible matrices $W$, the image vector $\Psi$ can be obtained through matrix inversion $\Psi = W^{-1}\Phi$, which finally has to be reshaped into a 2D matrix of pixel values. Futher, the matrix equation $\Phi = W\Psi$ represents the image being expanded in some basis given by $W$. For this study we mainly use Hadamard matrices as our basis expansion, ie. $W$ is a Hadamard matrix of order $N$ [18s]. A Hadamard matrix $H_n$ is defined as an $n\times n$ matrix of +1s and -1s with the property that the scalar product between any two distinct rows is 0 ie. each row is orthogonal to every other one. Thus $H_n$ must satisfy:

$$H_n H_n^T = H_n^T H_n = nI_n \tag{3}$$

where $H_n^T$ is the transpose of $H_n$. This is a property that allows for easy image reconstruction as it is easy to see that $H_n^{-1} = H_n^T/n$. A more serious reason to construct masks from Hadamard matrices is that this basis minimizes the mean square error of each pixel in our image [18s].

We have opaque masks that either block or transmit light ie. $W$ contains only values of 1s and 0s and thus is not a Hadamard matrix. However as outlined in [23s], it is possible to obtain a fully orthogonal measurement matrix with such a system. Consider the $H_2$ matrix:

$$H_2 = \begin{bmatrix} 1 & 1 \\ 1 & -1 \end{bmatrix}. \tag{4}$$

The problem is that our measurement matrices can only have values of 1 and 0 corresponding to the mirrors being on or off respectively, but if we consider the following two matrices:

$$G = \begin{bmatrix} 1 & 1 \\ 1 & 0 \end{bmatrix}, \qquad V = \begin{bmatrix} 0 & 0 \\ 0 & 1 \end{bmatrix}, \tag{5}$$

it is easy to see that $H_2 = G - V$. Thus if we have two sets of measurement vectors each using one of the complementary sets of masks,

$$\Phi_1 = G\Psi, \qquad \Phi_2 = V\Psi, \tag{6}$$

then subtraction of the second set gives the desired encoding matrix. This doubles the number of measurements required. However, if the complementary negative mask is projected immediately after its positive counterpart, one can eliminate an unwanted source of noise, namely low frequency source oscillations. In supplementary fig. S3, we compare the difference between using encoding masks derived from Hadamard matrices with [1, -1] and [1, 0] values. Here we can see that the image constructed from [1, 0] measurement has some artifacts created from low frequency THz source oscillations (indicated by the arrows), whereas the [1, -1] measurement have eliminated most of this type of artifact.

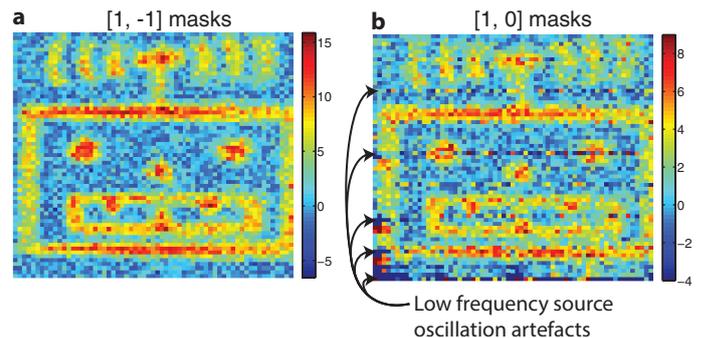

**Supplementary figure S3: [1, -1] vs [1, 0] masks.** Image obtained using Hadamard masks with values of [1, -1] in **a** and [1, 0] in **b**. Total number of measurements is 16384 for both pictures.

## §4 Scalar diffraction from two slits

As stated in the main text, the resolution of our imaging technique is limited by the thickness of our silicon photomodulator (115μm). To calculate the diffraction in our system, we follow the method outlined in ref [14s]. Therein, Kowarz solves the 2D Helmholtz equation for positive $z$-space with all scatterers, sources and diffracting apertures being located in negative $z$-space. His electric field solution $U(x,z)$ is the sum of two parts, a homogeneous propagating contribution $U_h(x,z)$ and an evanescent component $U_i(x,z)$:

$$U_h(x,z) = \int_{|u_x|\leq 1} A(u_x)e^{iku_x x}e^{ikz\sqrt{1-u_x^2}}\mathrm{d}u_x \qquad (7)$$

$$U_i(x,z) = \int_{|u_x|>1} A(u_x)e^{iku_x x}e^{-kz\sqrt{u_x^2-1}}\mathrm{d}u_x \qquad (8)$$

where $k$ is the free space wavenumber, $u_x$ is the directional wavevector in $x$ and $A(u_x)$ is a spectral amplitude function that is the Fourier transform of scatterer's field distribution in the plane $z=0$, ie.

$$A(u_x) = \frac{k}{2\pi}\int_{-\infty}^{\infty} U(x,0)e^{-iku_x x}\mathrm{d}x. \qquad (9)$$

This notation is known as the angular spectrum representation [15s]. We calculate the diffracted intensity distribution generated by two parallel slits with a field distribution given by

$$U(x,0) = K\left( \mathrm{rect}\left( \frac{x-a/2}{d} \right) + \mathrm{rect}\left( \frac{x+a/2}{d} \right) \right) \quad (10)$$

where $\mathrm{rect}$ is the rectangle function [16s], $d$ is the slits width, $a$ is the slits' center to center separation and $K$ is the incident wave amplitude. We are interested in the intensity distribution, defined as $I(x,z)\equiv|U(x,z)|^2=|U_h(x,z)+U_i(x,z)|^2$, for various slit separations. To model our system more accurately, we sum all the frequency contributions of our pulses, with each frequency component weighted by our pulse spectrum (Fig 2**b**) and a wavelength corresponding to our silicon dielectric.

In Fig. S4 we plot the intensity distribution, on a parallel plane at a distance equal to our modulator thickness (115μm), from two 20μm slits for various slit separations. For separations <65μm, the diffraction pattern is similar to that of a single slit. As the separation increases, the diffraction maxima arising from each slit become distinguishable. The white dashed line indicates the separation resolvable by the Rayleigh resolution criterion [17s], corresponding to a resolution of ~95μm which is well in agreement with our experimental estimate of 103(±7)μm.

## §5 Signal with increasing number of pixels

Here we investigate how experimental noise affects the three different masking schemes, outlined in the main text, as the number of pixels in the image is increased. For this, we take images under identical conditions (one after the other) of the circuit board in Fig 3**a** with increasing number of pixels. Our results are shown in supplementary Fig. S5. As shown in the main text, Hadamard masks have the most superior signal to noise followed by random masks and then by raster scanning. This is true for all image sizes. Raster scanning is most affected by detector noise due to the small signals emanating from a single aperture. On decreasing the aperture size, and increasing the number of pixels, image noise clearly increases. This effect is less significant for the multi-pixel approaches as these have larger associated signals. While multi-pixel patterns clearly have the benefit of increased signal to noise, the continual increase in the number of pixels leads to increased image noise, even for Hadamard imaging. This is because the signal from each individual pixel decreases as the number of pixels increase, and even though Hadamard matrices minimize the mean square error in each image pixel [18s] they do not completely remove all noise. One should also note that we have noise in our THz source which further degrades image quality as the number of measurements required to form the image increases. Interestingly, random masks seem to fair best as the number of pixels increases. It can easily be shown that it is an artefact caused by the simple reconstruction algorithm employed.

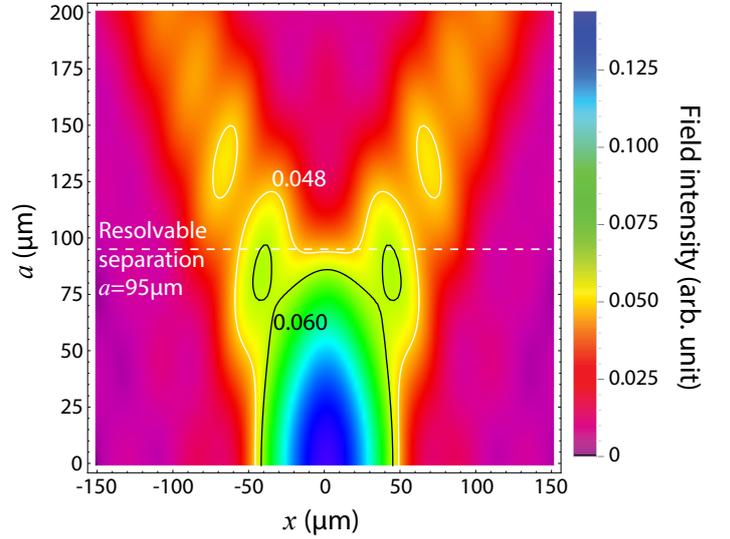

**Supplementary figure S4: Diffraction from two slits.** Intensity distribution (horizontal axis) from a slits as they separated apart (vertical axis). We plot the sum total intensity from an ensemble of spectrally weighted components that constitute our pulses (refer to experimental section for pulse spectrum). $d$=20μm, $z$=115μm in calculation.

## §6 Total variation minimization reconstruction

As stated in the main paper, we reconstruct our random mask images via a very simple algorithm where we sum the random masks with each one weighted by the detector readout for that mask. We use this simple algorithm due to its quick calculation times (~100ms) and for the Hadamard case it recovers the exact solution. However, the idea of using masks constructed from random matrices originates from compressed sensing [25s, 26s, 27s] where one usually minimizes the L1-norm of the vector (image in our case) that one wishes to sample in order to recover the correct solution. To obtain our images, we perform a total variation (TV) minimization. In other

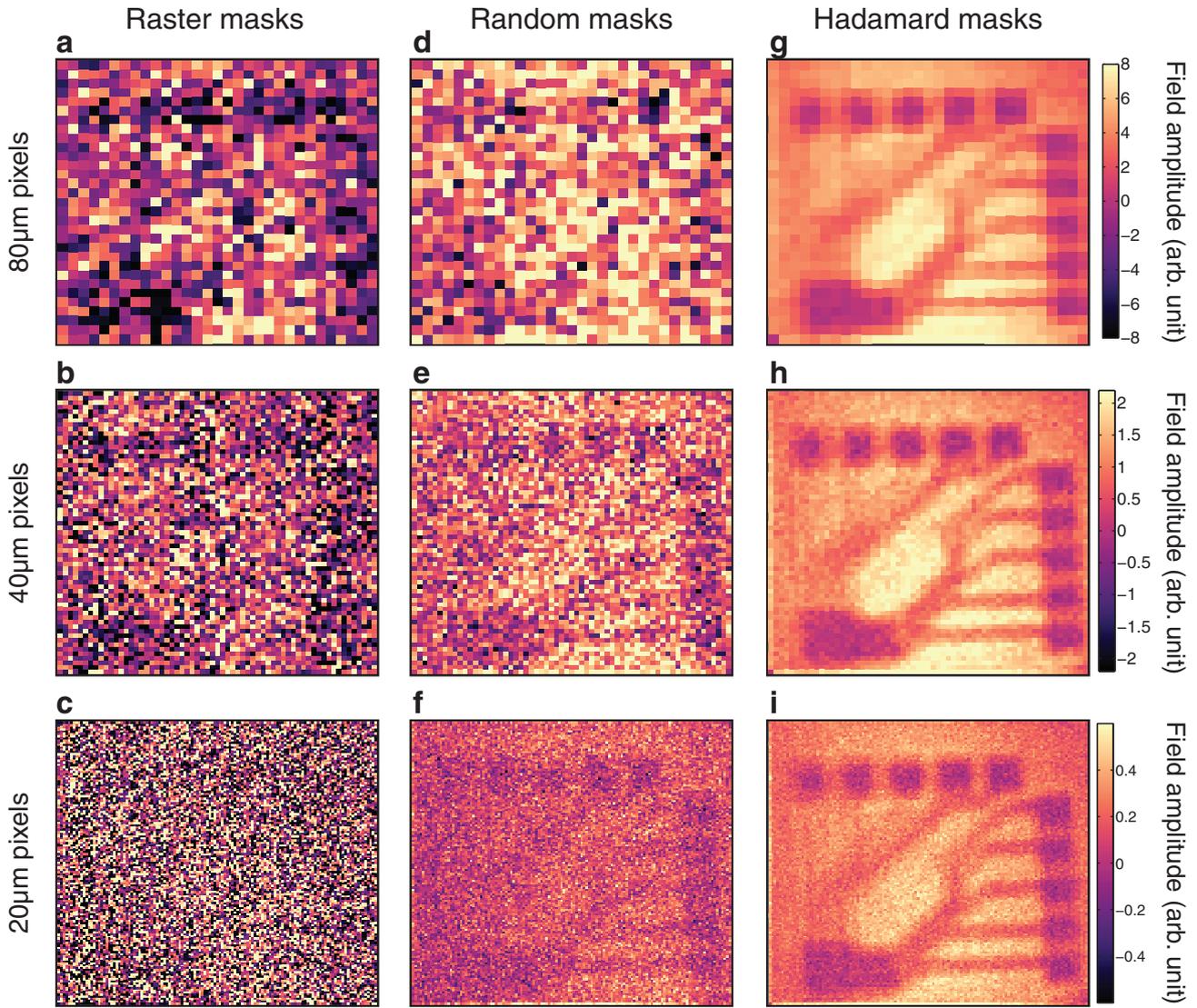

**Supplementary figure S5: Increasing image size. a-c,** Images obtained using raster masks with increasing number of pixels from 32×32 to 64×64 and 128×128, respectively. **d-f (g-i),** Images obtained using random (Hadamard) masks as number the of pixels is increased from 32×32 to 64×64 and 128×128, respectively. The vertical lines seen in part **c** are associated with periodic changes in lab environment. Note **a, b, & c** have been scaled by 0.9, 0.25 & 0.1, respectively, so as to be plotted on the same scale as all other images.

words, the mathematical problem is stated as

$$\text{minimize} \quad \text{TV}(\Psi)$$
$$\text{subject to} \quad \left\| W\Psi - \Phi \right\|_2 \leq \gamma, \quad (11)$$

where $W$ is our random measurement matrix, $\Phi$ is our vector of measurements, $\Psi$ is the image we are interested in, $\gamma$ is a variation relaxation parameter allowing us to determine how smooth the final image is and $\text{TV}$ is the total variation of a 2D image defined as

$$\text{TV}(x) \equiv \sum_i \sqrt{(\mathrm{D_h}x)_i^2 + (\mathrm{D_v}x)_i^2}, \quad (12)$$

where $x$ is a 2D image and $\mathrm{D_{h,v}}$ are the discretized gradient operators along the horizontal and vertical directions respectively. Our calculations were performed in Matlab 2013b using the L1-magic package [28s].

This algorithm is more complicated (taking us ~100s), however with it we obtain an image that has a significantly better signal to noise ratio as shown in supplementary Fig. S6. Here, we see that Hadamard is still superior. This is partly due to the fact that Hadamard matrices minimize the mean square error in each image pixel [18s] and partly due to the value of our relaxation parameter. In other words, we could further improve the quality of our random mask image with more careful considerations of our value for $\gamma$.

## §7 Image Filtering

In supplementary Fig. S7 we show THz images of certain sections of the manufactured circuit board (see Fig. 3**a** for design). With these images we show that it is possible to observe, using the strong polarization effects shown in figure 4, highly subwavelength (8µm) breaks along the conducting wires. Upon close investigation of Fig. S7, one can see the breaks manifesting as small localized increases in field amplitude. However, due to the noise embedded in the measurement this is not immediately obvious.

The experimental errors were minimized and a total of ~2650 pulses were utilized for each measurement, which results in rather noisey images (Fig S7) . Hence we are left to reduce the noise in our images via post-processing. For this, we employ a spatial Fourier filter and a spatial curvature denoising algorithm (outlined below). For Fourier filtering we use a Gaussian lowpass filter of size 15 with a standard deviation of 1.1, as implemented by the 'fspecial' command in MATLAB [19s]. These parameters were subjectively chosen based on subjective image quality. In contrast, the denoising algorithm, as outlined below and in Ref. [20s], has no subjective input.

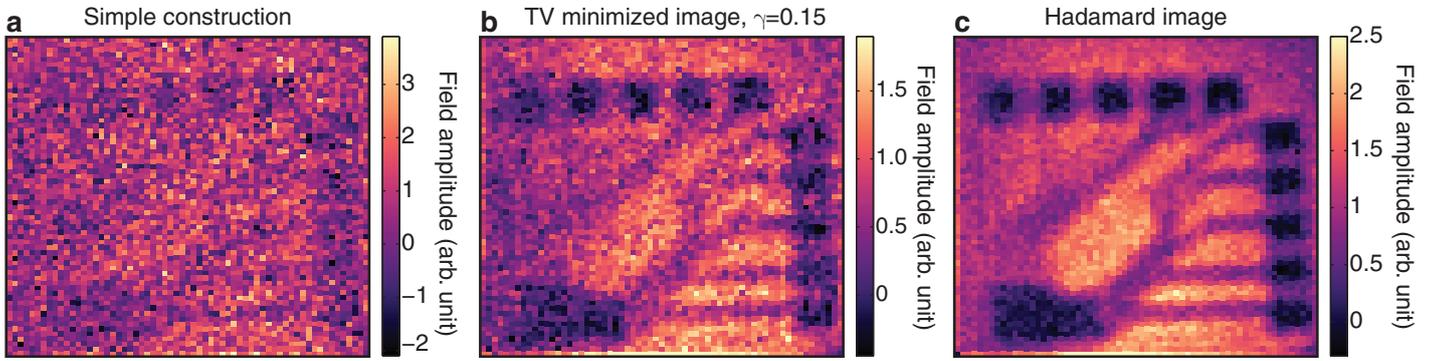

**Supplementary figure S6: Total variation minimized images.** 64×64 images of a circuit board where **a** and **b** have been obtained using random masks except **a** was constructed using our simple algorithm and **b** has bee constructed via a total variation minimization algorithm from the same data. **c** image of the same circuit board obtained via Hadamard masks.

Our noise is embedded within the spatial-curvature of our images. If we minimize the spatial-curvature, then the noise will also be minimized. However, we do not want to remove image features not due to noise, thus we have to put a constraint to limit the minimization. The constraint should also be chosen to reflect the nature of the noise to be filtered; Gaussian noise in our case. For this reason, we look at the square of the difference between our denoised and original images. The denoised image is obtained by minimization of its cost function, $C$, given by

$$C = \frac{1}{N} \sum_{i=1}^{N} \left( \frac{\sum_{j=1}^{N} (w_{ij}\psi_j) - \phi_i}{\sigma_s} \right)^2 + \lambda \left( \left| \frac{d^2\Psi'}{dx^2} \right| + \left| \frac{d^2\Psi'}{dy^2} \right| \right) (13)$$

where $\Psi'$ is the image vector expressed in 2D format, $\sigma_s$ is the standard deviation of the noise in the measurement of $\phi_i$. The first term in eq. (13) represents a $\chi^2/N$ distribution of the image with respect to the measured data, and the second term represents the total spatial-curvature of the image. $\lambda$ is the regularization parameter dictating the level of smoothing: larger values lead to smoother denoised images. The algorithm starts with a value of $\lambda=1$ and automatically increases this value to ensure that, once optimized, $\chi^2/N \approx 1$.

In supplementary figure S8 we show the raw images in supplementary Fig. S7 filtered by the two methods outlined above. Both methods considerably improve the looks of the images, and more importantly they preserve the features we wish to show: the breaks manifesting as small localized increases in amplitude.

### Supplementary References

## Unfiltered break images

### Horizontal polarization

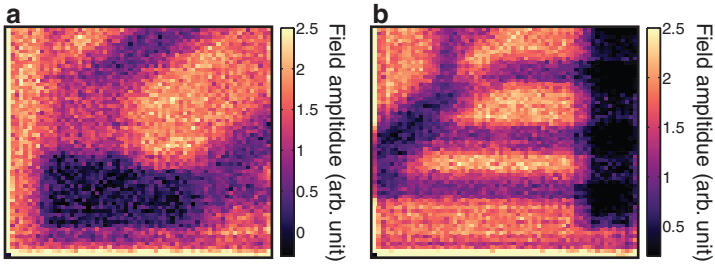

### Vertical polarization

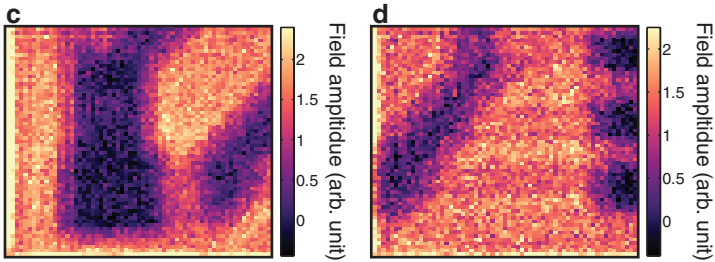

**Supplementary figure S7: Unfiltered images. a-b,** Image of Break A and B, respectively, with horizontal polarization. **c-d,** Image of Break A and B, respectively, with vertical polarization. Number of pixels is 64×64 with each pixel being 20µm in size for all images.

## Spatial Fourier filtered break images

### Horizontal polarization

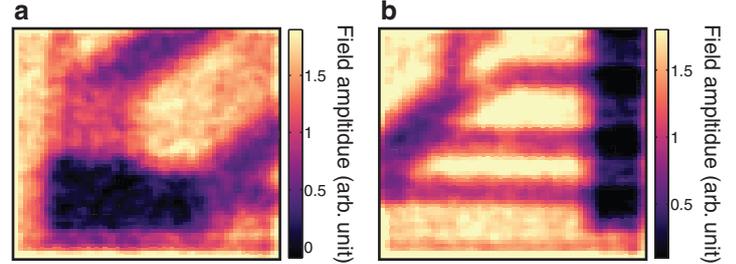

### Vertical polarization

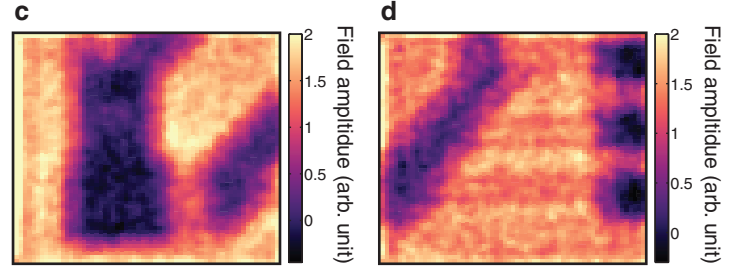

## Denoised break images

### Horizontal polarization

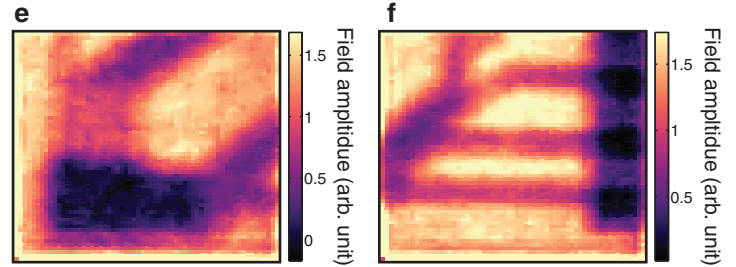

### Vertical polarization

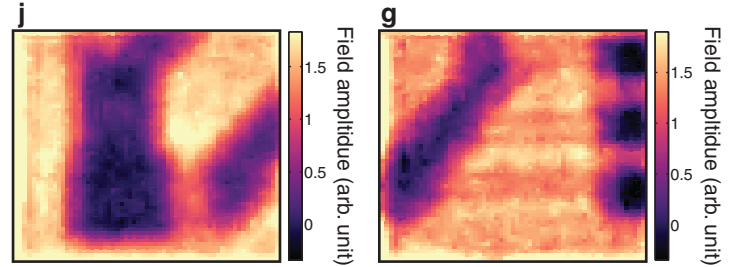

**Supplementary figure S8: Filtered images. a-d,** Fourier filtered images of Breaks A and B from supplementary figure S4. **e-g,** Denoised images of Breaks A and B from supplementary figure S4.